\documentclass[aps,preprint]{revtex4}%
\usepackage{amsfonts}
\usepackage{amsmath}
\usepackage{amssymb}
\usepackage{graphicx}
\usepackage{epstopdf}%
\providecommand{\U}[1]{\protect\rule{.1in}{.1in}}

\begin{document}
\title{Thermal fluctuations and dynamic modelling of a dAFM cantilever}
\author{Elena Pierro$^{1}$, Francesco Bottiglione$^{2}$, Giuseppe Carbone$^{2,3,4,5}$}
\affiliation{$^{1}$Scuola di Ingegneria, Universit\`{a} degli Studi della Basilicata, 85100
Potenza, Italy}
\affiliation{$^{2}$Department of Mechanics Mathematics and Management - Politecnico di
Bari, v.le Japigia 182, 70126 Bari, Italy}
\affiliation{$^{3}$Physics Department M. Merlin, CNR Institute for Photonics and
Nanotechnologies U.O.S. Bari via Amendola 173, 70126 Bari, Italy}
\affiliation{$^{4}$Department of Mechanical Engineering, Imperial College London, South
Kensington Campus, Exhibition Road London SW7 2AZ, United Kingdom}
\affiliation{$^{5}$Center for Nonlinear Science, University of North Texas, P. O. Box
311427, Denton, Texas 76203-1427, USA.}
\begin{abstract}
We discuss the Brownian thermal noise which affects the cantilever dynamics of
a dAFM (dynamic atomic force microscope), both when it works in air and in
presence of water. Our scope is to accurately describe the cantilever
dynamics, and to get this result we deeply investigate the relationship
between the cantilever thermal fluctuations and its interactions with the
surrounding liquid. We present a relatively simple and very easy-to-use
analytical model to describe the interaction forces between the liquid and the
cantilever. The novelty of this approach is that, under the assumption of
small cantilever oscillations, by using the superposition principle we found a
very simple integral expression to describe fluid-cantilever interactions.
More specifically we note that, beside including fluid inertia and viscosity
(which is common to many existing models in the literature) an additional
diffisivity term needs to be considered, whose crucial influence for the
correct evaluation of the cantilever response to the thermal excitation is
shown in the present paper. The coefficients of our model are obtained by
using numerical results for a 2D fluid flow around a vibrating rectangular
cross-section, and depend on the distance from the wall. This allowed us to
completely characterize the dynamics of a dAFM cantilever also when it
operates in tilted conditions. We validate the analytical model by comparing
our results with numerical and experimental dAFM data previously presented in
literature, and with experiments carried out by ourselves. We show that we can
provide extremely accurate prediction of the beam response up and beyond the
second resonant peak.

\end{abstract}
\maketitle

\part{Introduction}

The dynamic atomic force microscope (dAFM) consists of an oscillating
microcantilever which holds a sharp nanoscale tip that intermittently
interacts with the sample, close to its first resonance frequency. There are
many fields of applications of dAFM, and ranges from measuring topography of
organic and inorganic materials at nanometer length scales, to the accurate
quantification of sample properties in materials science. Despite many years
of investigations on dAFM dynamics \cite{1}, several aspects still demand
further attention from the research community. Because of the complexity
related to the small scales involved, where different physical effects coexist
together, it is a very common approach to study each phenomenon separately,
and to draw conclusions from each analysis \cite{2,3}. This is often a very
good way to clarify aspects completely unknown before, but for more detailed
insights, it is also necessary to analyze different effects together. This is
the case of the dAFM\ research field, in which, in particular, it is extremely
important to analyse both the tip-sample and the fluid-structure interactions.
For such instruments a clear signal is one of the utmost requirements in order
to extract correct information from the measurement at so small length scales.
However this is not a so straightforward task to be achieved because of the
forces which act on the cantilever in operational conditions, especially when
a liquid environment is required, e.g. in physiological buffer solutions
\cite{4,5,6,7}. In such case, in particular, besides the tip-sample
interaction, the presence of a liquid deeply modifies the response spectrum
and is the origin of the so called Brownian thermal noise. Indeed, even though
the molecular size of the fluid molecules is negligibly small in comparison to
the size of the cantilever, the dynamics of the cantilever itself is strongly
dependent on Brownian fluctuations, since the same molecular processes are
responsible for both the dissipation and the fluctuations, due to the
collisions between the fluid molecules and the cantilever. But to achieve a
good AFM resolution, noise sources must be reduced as much as possible, since
the lowest signal level that can be considered as a valid information must be
not mixed up with external noise. In spite of this, the cantilever thermal
motion is also utilized to calibrate the cantilever's spring constant \cite{8}
and to extract the resonance frequencies and quality factors of its resonances
as well \cite{9}. Therefore, a good understanding and theoretical modelling of
this effect related to the equilibrium of the cantilever with the surrounding
liquid is of paramount importance, whatever is the scope, i.e. either to study
ad hoc control system to separate the signal from noise, or to obtain useful
insights about the cantilever properties. With such an aim, several studies
have been presented up to now \cite{10,11,12}, but the main difficulty for a
proper description of the cantilever response to the thermal fluctuations is
the fluid-structure interaction modelling. In order to overcome the complexity
of the problem, some assumptions have been made in the previous studies, such
as those of neglecting inertia \cite{13} and the diffusivity
\cite{14,15,Paul2006,Cole2007} terms, which, instead, become important from
intermediate to high frequency range. In this paper the authors present an
analytical model to describe the drag of the liquid on the cantilever which
takes into account, in a relatively simple way, all the three mentioned
contributions: (i) the fluid viscosity, (ii) fluid inertia, and (iii) the
fluid vorticity diffusivity. It is shown, in particular, that the correct
evaluation of the cantilever response to the thermal excitation cannot neglect
the above mentioned effects. Moreover, experimental results of an AFM
cantilever working in air are presented, which definitively assess the
accuracy of the model.

The paper is organized as follows: in Sec.I, the model used to describe the
cantilever dynamics is presented; in Sec.II the new general expression of the
liquid drag is heuristically derived; in Sec.III the relation between the
cantilever response and the Brownian loading spectra is shown; in Sec. IV our
analytical model is compared with previous works and with our experimental
results; in Sec. V we provide concluding remarks; in Appendix A we show how to
calculate the susceptibility function of the cantilever; in Appendix B the
Fluctuation Dissipation Theorem for a 1DOF system is shown;\ in Appendix C we
analytically derive the Fluctuation Dissipation Theorem for the continuous
system we are studying, i.e. for the cantilever;\ in Appendix D the
Equiripartition theorem for the cantilever case is derived.

\section{Cantilever dynamics}

We study the dynamics of a cantilever immersed in a viscous fluid (see Figure
\ref{Figure 1}) near a rigid flat wall, with $L$, $B$, and $H$ respectively
the length, the width and the thickness of the rectangular cross section. We
will assume that $L\gg B\gg H$, as well as that the transversal displacement
$\left\vert w\left(  x,t\right)  \right\vert \ll L$. This enables us to use
the Bernoulli theory of transversal vibrations and therefore neglect the
influence of shear stress in the beam. The general motion equation of the
cantilever is therefore: \begin{figure}[ptb]
\begin{center}
\includegraphics[
height=9cm,
]{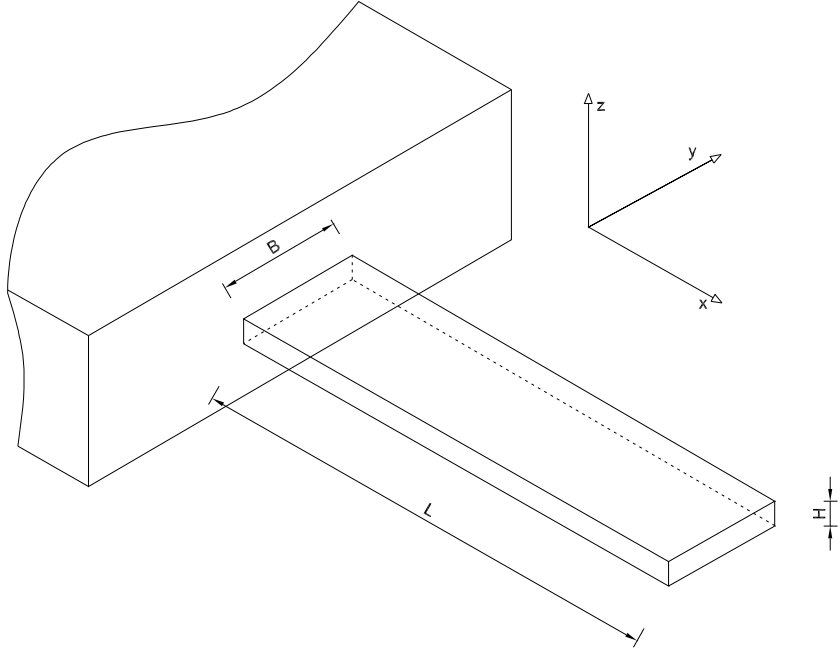}
\end{center}
\caption{dAFM microcantilever.}%
\label{Figure 1}%
\end{figure}%
\begin{equation}
EI\frac{\partial^{4}w\left(  x,t\right)  }{\partial x^{4}}+\rho A\frac
{\partial^{2}w\left(  x,t\right)  }{\partial t^{2}}=f_{B}\left(  x,t\right)
+f_{L}\left(  x,t\right)  \label{1}%
\end{equation}
where $\rho$ is the bulk density of the material the cantilever is made of,
$E$ is the Young'modulus and $A$ is the area of the cross section of the beam,
i.e. $A=BH$. In the RHS of the above Eq. (\ref{1}) we have introduced two
terms. The first term $f_{B}\left(  x,t\right)  $ is the chaotic force per
unit length acting on the beam as a consequence of the thermal fluctuation of
the molecules constituting the liquid: we will refer to this type of
excitation term as the Brownian force.

The term $f_{L}\left(  x,t\right)  $ is the force per unit length which the
liquid, in a continuum sense, exerts on the cantilever, which is usually
calculated by means of complex numerical simulations. However, under the
hypothesis of small transversal displacements $\left\vert w\left(  x,t\right)
\right\vert \ll B$, we can assume that the response of the fluid is linear.
Under this condition the force $f_{L}\left(  x,t\right)  $ can be expressed in
terms of the response function $G\left(  x,x^{\prime},t\right)  $:%
\begin{equation}
f_{L}\left(  x,t\right)  =\int_{-\infty}^{t}d\tau\int_{0}^{L}dx^{\prime
}G\left(  x,x^{\prime},t-\tau\right)  \frac{\partial v\left(  x^{\prime}%
,\tau\right)  }{\partial\tau}\label{2}%
\end{equation}
where $v\left(  x,\tau\right)  $ is the transversal velocity of the generic
cantilever section, i.e. $v\left(  x,\tau\right)  =w_{t}\left(  x,t\right)  $
, where the subscript $t$ denotes the partial time-derivative. Moreover,
causality implies the linear response function of the liquid must vanish for
$t<0$. This also implies that the quantity%
\begin{equation}
G\left(  x,x^{\prime},\omega\right)  =\int_{-\infty}^{+\infty}dtG\left(
x,x^{\prime},t\right)  \exp\left(  -i\omega t\right)  \label{3}%
\end{equation}
must be analytic in lower half-plane of the complex domain, and therefore must
satisfy the Kramer-Kronig relations \cite{20}. However, measurements or
numerical calculations contain always some errors or approximations that do
not guarantee the Kramer-Kroning relations to be exactly satisfied. This
restriction in general leads to some difficulties in the calculations of the
linear response function $G\left(  x,x^{\prime},t\right)  $ of the liquid from
the measured or numerically calculated frequency response, i.e. it is often
not possible to determine the quantity $G\left(  x,x^{\prime},t\right)  $ by
simply taking the inverse Fourier transform of the measured frequency response
of the system. However the task is much more simplified if the analytical form
of this function is already known. For this reason, the authors have derived a
new analytical model of the liquid response $G\left(  x,x^{\prime},t\right)
$, that will be presented in the next section.

\section{Fluid-cantilever interaction}

We first observe that some hypothesis about the time dependence of the liquid
response function $G\left(  t\right)  $ must be fulfilled. Indeed, the
assumption that $L\gg B\gg H$ allows us to consider that the response of the
fluid depends only on local quantities, i.e. the fluid motion is locally two
dimensional \cite{13,15}. In such case we can write $G\left(  x,x^{\prime
},t\right)  =G\left(  x,t\right)  \delta\left(  x^{\prime}-x\right)  $. Now,
the leading-order incompressible flow generated by an isolated body
oscillating at small amplitude is governed by the unsteady Stokes equation%
\begin{align}
\rho_{L}\frac{\partial\mathbf{v}}{\partial t}+\rho_{L}\left(  \mathbf{v}%
\cdot\nabla\right)  \mathbf{v}  &  =-\nabla p+\eta\nabla^{2}\mathbf{v}%
\label{4}\\
\nabla\cdot\mathbf{v}  &  =0\nonumber
\end{align}
where $\mathbf{v}$ is the velocity of the fluid, $\eta$ is the liquid
viscosity, $\rho_{L}$ the liquid density. Observe that the non-linear term
$\rho_{L}\left(  \mathbf{v}\cdot\nabla\right)  \mathbf{v}$ can be neglected in
comparison with the time derivative of Eq.\ref{4}. Indeed, by estimating the
two terms, we obtain that $\rho_{L}\left(  \mathbf{v}\cdot\nabla\right)
\mathbf{v\sim}\left(  w_{0}\omega\right)  ^{2}/B$, and the time derivative
$\rho_{L}\partial\mathbf{v/}\partial t\mathbf{\sim}w_{0}\omega^{2}$, being $B$
the characteristic dimensions of the moving bodies, i.e. the width of the
cantilever cross section, and $w_{0}$ the amplitude of the oscillations.
Therefore the non linear convective term can be neglected if $w_{0}\ll B$,
which is indeed one of the hypotheses of our model. Moreover, Eq. (\ref{4})
can be rephrased in terms of the vorticity $\mathbf{W}\mathbf{=}\nabla
\times\mathbf{v}$ as $\rho_{L}\left(  \partial\mathbf{W/}\partial t\right)
=\eta\nabla^{2}\mathbf{W}$, showing that the vorticity is governed by a
diffusive-like equation which generates an exponential decay of the velocity
as we move far from the walls of the bodies towards the interior of the
liquid. The exponential decay allows to estimate the thickness $h$ of the
layer of fluid within which the flow is rotational and velocity diffusion is
important. It is known \cite{16} that $h\sim\omega^{-1/2}$ where $\omega$ is
the characteristic frequency of the motion. Out of the layer the term
$\nabla^{2}\mathbf{v}$ can be neglected in Eq. \ref{4} so that we have
potential flow. Since the value of $h$ depends on the frequency $\omega$, it
follows that at large frequencies the thickness $h$ will be very small and the
response of the fluid will be then governed by inertia effects, i.e. it will
depend proportionally on the acceleration $v_{t}\left(  x,t\right)  $ of the
moving bodies. In this limiting case $G\left(  x,t\right)  =-\mu\left(
x\right)  \delta\left(  t\right)  $, where $\delta\left(  t\right)  $ is the
Dirac delta function. If, on the contrary, the motion is sufficiently slow
(low $\omega$-values) then the term $\partial\mathbf{v/}\partial t$ can be
neglected and the quantity $h$ becomes larger than the characteristic
dimensions of the moving bodies. The motion of the fluid becomes steady and
the response of the fluid should depend linearly on the velocity $v$ of the
moving bodies, i.e. for $t\geq0$, we have $G\left(  x,t\right)  =-c\left(
x\right)  $. At intermediate frequency the effect of diffusivity should be
very important. It is known \cite{16} that for the in-plane motion of flat
plate the motion of the liquid is governed by a pure diffusive equation, and
in this case $G\left(  x,t\right)  $ takes the form $G\left(  x,t\right)
=-\alpha\left(  x\right)  t^{-1/2}$ for $t\geq0$, which represents the
contribution to the fluid force coming from the diffusion of the tangential
velocity of the plate into the interior of the liquid. Since Eqs. (\ref{4})
are linear, the force the liquid exerts on the body will be given by the sum
of the three contributions and we can heuristically give a general expression
for $G\left(  x,t\right)  $ as%
\begin{align}
G\left(  x,t\right)   &  =-c\left(  x\right)  -\frac{\alpha\left(  x\right)
}{\sqrt{t}}-\mu\left(  x\right)  \delta\left(  t\right)  ;\qquad
t\geq0\label{G_function}\\
G\left(  x,t\right)   &  =0;\qquad t<0\nonumber
\end{align}
where $c\left(  x\right)  $ is the damping, $\alpha\left(  x\right)  $ is the
diffusive coefficient, and $\mu\left(  x\right)  $ is the inertia term.
Eq.(\ref{G_function}), therefore, satisfies the causality principle. We
observe, in particular, that for the case of a sphere moving in a liquid, the
response of the fluid has exactly the form given by Eq. (\ref{G_function}),
see for example \cite{16}. Moreover, by substituting Eq.(\ref{G_function}) and
Eq.(\ref{2}) in Eq.(\ref{1}), we get%
\begin{align}
&  EI\frac{\partial^{4}w\left(  x,t\right)  }{\partial x^{4}}+\left[  \rho
A+\mu\left(  x\right)  \right]  \frac{\partial^{2}w\left(  x,t\right)
}{\partial t^{2}}+c\left(  x\right)  \frac{\partial w\left(  x,t\right)
}{\partial t}+\label{complete_equation}\\
+\alpha\left(  x\right)  \int_{-\infty}^{t}d\tau\frac{1}{\sqrt{t-\tau}}%
\frac{\partial^{2}w\left(  x,\tau\right)  }{\partial\tau^{2}}  &
=f_{B}\left(  x,t\right) \nonumber
\end{align}
and in particular, the linear response function $\chi\left(  x,x^{\prime
},t\right)  $ of this equation, also referred to as susceptibility, can be
obtained by solving the fundamental problem (see Appendix A):%
\begin{align}
&  EI\frac{\partial^{4}\chi\left(  x,x^{\prime},t\right)  }{\partial x^{4}%
}+\left[  \rho A+\mu\left(  x\right)  \right]  \frac{\partial^{2}\chi\left(
x,x^{\prime},t\right)  }{\partial t^{2}}+c\left(  x\right)  \frac{\partial
\chi\left(  x,x^{\prime},t\right)  }{\partial t}+\label{fundamental problem}\\
+\alpha\left(  x\right)  \int_{-\infty}^{t}d\tau\frac{1}{\sqrt{t-\tau}}%
\frac{\partial^{2}\chi\left(  x,\tau\right)  }{\partial\tau^{2}}  &
=\delta\left(  x-x^{\prime}\right)  \delta\left(  t\right) \nonumber
\end{align}

Eq. (\ref{G_function}) is, of course, approximated, but in what follows it
will be shown that it works very well. In Eq. (\ref{G_function}) the
$x$-dependence can be present only as a consequence of the $x$- dependent
distance of the cantilever cross sections from the wall, i.e. only when the
beam is tilted, which is a case also covered by this study. Once the
analytical form of the response function is known, the three coefficients can
be determined by finding the best fitting with the existing accurate
computational fluid dynamics non-dimensional solutions of a 2D fluid flowing
around a vibrating rectangular cross-section, as reported in Ref. \cite{15}.
In Table1, the three coefficients are shown, calculated in correspondence of
the cantilever free end $x=L$, for different values of $\Delta=2d/B$, where
$d$ is the wall distance. Now, Fourier transforming Eq.
(\ref{complete_equation}) we obtain%
\begin{equation}
\frac{\partial^{4}w\left(  x,\omega\right)  }{\partial x^{4}}-B\left(
x,\omega\right)  ^{4}w\left(  x,\omega\right)  =f_{B}\left(  x,\omega\right)
\label{7}%
\end{equation}
where we define the function $B\left(  x,\omega\right)  $
\begin{equation}
B\left(  x,\omega\right)  =\sqrt[4]{-\frac{i\omega c\left(  x\right)  -\left[
\rho A+\mu\right]  \omega^{2}+i\omega~\alpha\left(  x\right)  \sqrt{i\omega
\pi}}{EI}} \label{8}%
\end{equation}

\begin{center}%
\begin{tabular}
[c]{|c|c|c|c|}\hline
$\Delta$ & $c\ \mathrm{[kg\ s}^{-1}\mathrm{]}$ & $\alpha\ \mathrm{[kg\ s}%
^{-1/2}\mathrm{]}$ & $\mu\ \mathrm{[kg]}$\\\hline
$0.5$ & $0.157$ & $8.32\times10^{-6}$ & $5.73\times10^{-7}$\\\hline
$1$ & $0.042$ & $1.25\times10^{-5}$ & $4.65\times10^{-7}$\\\hline
$2$ & $0.016$ & $2.13\times10^{-5}$ & $4.13\times10^{-7}$\\\hline
$4$ & $8.6\times10^{-3}$ & $3.72\times10^{-5}$ & $3.81\times10^{-7}$\\\hline
$6$ & $6.05\times10^{-3}$ & $4.91\times10^{-5}$ & $3.65\times10^{-7}$\\\hline
$8$ & $4.91\times10^{-3}$ & $5.65\times10^{-5}$ & $3.56\times10^{-7}$\\\hline
$10$ & $4.31\times10^{-3}$ & $6.1\times10^{-5}$ & $3.51\times10^{-7}$\\\hline
\end{tabular}

{\small Tab.1 - The three coefficients of the fluid response, at the free end
of the cantilever }$x=L${\small , for different values of the distance
}$\Delta=2d/B$ {\small of the cantilever from the substrate.}
\begin{figure}[ptb]
\begin{center}
\includegraphics[
height=9cm
]{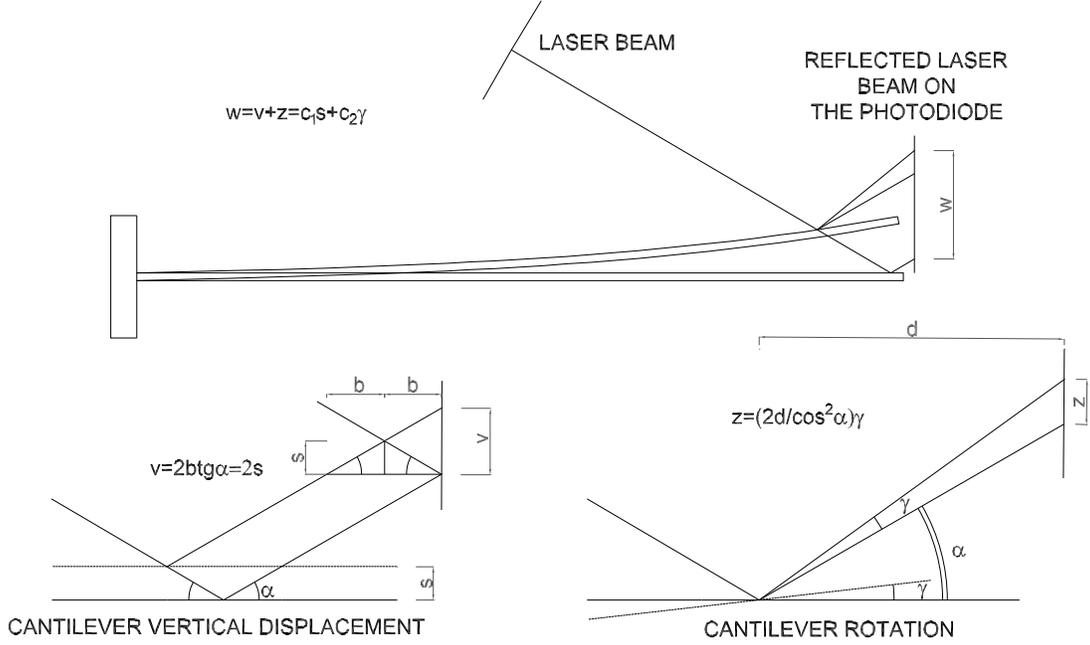}
\end{center}
\caption{The total displacement $w_{tot}$ of the laser beam on the photodiode
is given by the weighted sum of both the vertical displacement $s\left(
x,t\right)  $ and the rotation $\gamma\left(  x,t\right)  $ of the
cantilever.}%
\label{Laser_displacement_contributes}%
\end{figure}
\end{center}

\section{The fluctuation dissipation theorem}

In AFM systems the micrometer size of the cantilever makes it particularly
sensitive to Brownian fluctuations so that the thermal induced fluctuations of
the cantilever displacement cannot be neglected, thus making the measurement
of the response function of the cantilever through its thermal fluctuations a
viable technique. This is done by employing the Fluctuation Dissipation
Theorem (FDT) (see Appendix B). The FDT states that the susceptibility
function $\chi\left(  x_{1},x_{2},t\right)  $ of the cantilever, which is the
displacement $w\left(  x_{1},\omega\right)  $ at point $x_{1}$ due to the
action of a concentrated unit force $F$ at point $x_{2}$, is proportional to
the time-derivative of the correlation function of the thermal fluctuations of
the cantilever displacements (see Appendix C) \cite{17}, that is%
\begin{equation}
\chi\left(  x_{1},x_{2},t\right)  =-\beta H\left(  t\right)  \frac{\partial
}{\partial t}\left\langle w\left(  x_{1},0\right)  w\left(  x_{2},t\right)
\right\rangle \label{9}%
\end{equation}
where $\beta=\left(  k_{B}T\right)  ^{-1}$ and $k_{B}$ is Boltzmann's
constant, $T$ the absolute temperature and $H\left(  t\right)  $ is the
Heaviside unit step function. The susceptibility function can be used to
rephrase the solution of Eq. (\ref{1}) as
\begin{equation}
w\left(  x_{1},t\right)  =\int_{-\infty}^{t}d\tau\int_{0}^{L}dx_{2}\chi\left(
x_{1},x_{2},t-\tau\right)  f_{B}\left(  x_{2},\tau\right)  \label{10}%
\end{equation}
Moving from Eq.(\ref{9}), the relation between the Cross Power Spectral
Density (CPSD) $R\left(  x_{1},x_{2},\omega\right)  =\int dt~\left\langle
w\left(  x_{1},0\right)  w\left(  x_{2},t\right)  \right\rangle \exp\left(
-i\omega t\right)  $ of the cantilever thermal fluctuations and the imaginary
part of the time-Fourier transform of the susceptibility function, i.e. the
imaginary part of the complex compliance $\chi\left(  x_{1},x_{2}%
,\omega\right)  =\int dt\chi\left(  x_{1},x_{2},t\right)  \exp\left(  -i\omega
t\right)  $, can be easily derived (see Appendix C) \cite{18} as%
\begin{equation}
R\left(  x_{1},x_{2},\omega\right)  =-\frac{2k_{B}T}{\omega}\operatorname{Im}%
\chi\left(  x_{1},x_{2},\omega\right)  \label{11}%
\end{equation}
Eq. (\ref{11}) will be used to calculate the PSD of the thermal oscillations
of the free end of the cantilever. We observe that the CPSD $R\left(
x_{1},x_{2},\omega\right)  $ is, in general, not $\omega$-independent (as it
should be in case of white noise), since, as we show later, the
velocity-diffusive and the inertia terms of the liquid response, do not make
$\operatorname{Im}[\chi\left(  x_{1},x_{2},\omega\right)  ]$ proportional to
the radian frequency $\omega$. Equation (\ref{11}) shows that, in order to
calculate the CPSD of the cantilever displacement fields, it is enough to
measure or calculate the response function $\chi\left(  x_{1},x_{2}%
,\omega\right)  $. However, this quantity is strongly affected by the presence
of the liquid. By solving Eq. (\ref{7}) it is possible to determine the
complex compliance of the cantilever for the transversal displacement and
therefore determine the CPSD $R\left(  x_{1},x_{2},\omega\right)  $. In order
to define the correct signal information which is actually read by the AFM
system, we point out that the angular rotation of the beam cross section gives
a huge contribution to signal read by the photodiode. In Fig.
\ref{Laser_displacement_contributes}, indeed, one can clearly observe that the
total displacement $w_{\mathrm{tot}}\left(  t\right)  $ of the reflected laser
spot on the photodiode surface can be calculated as the weighted sum of both
the vertical $s\left(  x,t\right)  =w\left(  x,t\right)  $ displacement and
the angular $\gamma\left(  x,t\right)  =\dot{s}\left(  x,t\right)  =$ $\left(
\partial w/\partial x\right)  _{l,t}$ rotation of the cantilever, i.e.%

\begin{equation}
w_{\mathrm{tot}}\left(  x,t\right)  =c_{1}s\left(  x,t\right)  +c_{2}%
\gamma\left(  x,t\right)  \label{total_displacement}%
\end{equation}
Therefore we will assume in what follows that the CPSD\ of the quantity
$w_{\mathrm{tot}}\left(  x,t\right)  $ is $S_{\mathrm{tot}}\left(  x_{1}%
,x_{2},\omega\right)  =\int dt\left\langle w_{\mathrm{tot}}\left(
x_{1},0\right)  w_{\mathrm{tot}}\left(  x_{2},t\right)  \right\rangle
\exp\left(  -i\omega t\right)  $. So we get, at the cantilever free end
$x_{1}=x_{2}=L$%
\begin{align}
\left\langle w_{\mathrm{tot}}\left(  L,0\right)  w_{\mathrm{tot}}\left(
L,t\right)  \right\rangle  &  =c_{1}^{2}\ \left\langle w\left(  L,0\right)
w\left(  L,t\right)  \right\rangle +\label{cross correlation}\\
&  +c_{1}c_{2}\left[  \left\langle w_{x}\left(  L,0\right)  w\left(
L,t\right)  \right\rangle +\left\langle w\left(  L,0\right)  w_{x}\left(
L,t\right)  \right\rangle \right]  +c_{2}^{2}\left\langle w_{x}\left(
L,0\right)  w_{x}\left(  L,t\right)  \right\rangle \nonumber
\end{align}
Now recall that%
\begin{equation}
\int dt\left\langle w\left(  L,0\right)  w\left(  L,t\right)  \right\rangle
e^{-i\omega t}=R\left(  L,L,\omega\right)  =-\frac{2k_{B}T}{\omega
}\operatorname{Im}\chi\left(  L,L,\omega\right)  \label{Def}%
\end{equation}
so taking the derivative we also obtain%
\begin{align}
\int dt\left\langle w_{x}\left(  L,0\right)  w\left(  L,t\right)
\right\rangle e^{-i\omega t}  &  =R_{1}\left(  L,L,\omega\right)
=-\frac{2k_{B}T}{\omega}\operatorname{Im}\chi_{1}\left(  L,L,\omega\right)
\label{Def_R}\\
\int dt\left\langle w\left(  L,0\right)  w_{x}\left(  L,t\right)
\right\rangle e^{-i\omega t}  &  =R_{2}\left(  L,L,\omega\right)
=-\frac{2k_{B}T}{\omega}\operatorname{Im}\chi_{2}\left(  L,L,\omega\right)
\nonumber\\
\int dt\left\langle w_{x}\left(  L,0\right)  w_{x}\left(  L,t\right)
\right\rangle e^{-i\omega t}  &  =R_{12}\left(  L,L,\omega\right)
=-\frac{2k_{B}T}{\omega}\operatorname{Im}\chi_{12}\left(  L,L,\omega\right)
\nonumber
\end{align}
where $\left(  \cdot\right)  _{i}=\partial\left(  \cdot\right)  /\partial
x_{i}$ and $\left(  \cdot\right)  _{ij}=\partial^{2}\left(  \cdot\right)
/\partial x_{i}\partial x_{j}$. The power spectrum of the signal recorded by
the photodiode is then%

\begin{align}
S\left(  \omega\right)   &  =S_{\mathrm{tot}}\left(  L,L,\omega\right)  =\int
dt\left\langle w_{\mathrm{tot}}\left(  L,0\right)  w_{\mathrm{tot}}\left(
L,t\right)  \right\rangle \exp\left(  -i\omega t\right) \nonumber\\
&  =c_{1}^{2}R\left(  L,L,\omega\right)  +c_{1}c_{2}\left[  R_{1}\left(
L,L,\omega\right)  +R_{2}\left(  L,L,\omega\right)  \right]  +c_{2}^{2}%
R_{12}\left(  L,L,\omega\right) \label{tot_CPSD}\\
&  =-\frac{2k_{B}T}{\omega}\operatorname{Im}\left\{  c_{1}^{2}\chi\left(
L,L,\omega\right)  +c_{1}c_{2}\left[  \chi_{1}\left(  L,L,\omega\right)
+\chi_{2}\left(  L,L,\omega\right)  \right]  +c_{2}^{2}\chi_{12}\left(
L,L,\omega\right)  \right\} \nonumber
\end{align}
\begin{figure}[ptb]
\begin{center}
\includegraphics[
height=5.77cm,
width=7.99cm
]{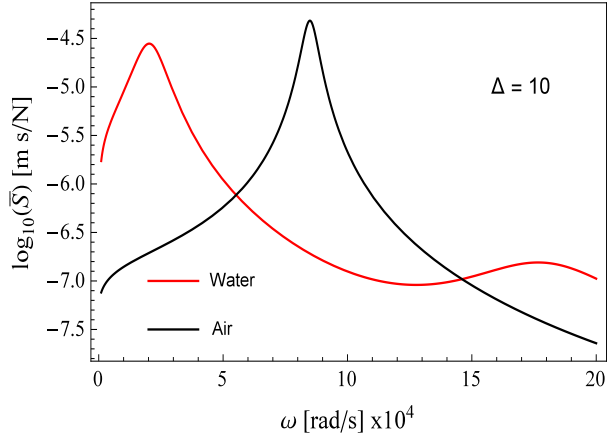}
\end{center}
\caption{The cantilever thermal power spectrum in water (red curve) and in air
(black curve), $\Delta=2d/B=10.$}%
\label{Figure 2}%
\end{figure}

\section{Results}

In this section we discuss the main results of our investigation. We compare
the thermal response of the cantilever tip in air and in water, show how
important is the influence of the diffusive-velocity term related to the
parameter $\alpha\left(  x\right)  $, and also discuss the influence of the
distance between the cantilever and the underlying substrate. The geometrical
and mechanical properties of the cantilever, we have investigated, are listed
in Tab.2. At first, we will focus on the PSD of the transversal displacement
of the free end of the cantilever, i.e. we will focus on the quantity
$S\left(  \omega\right)  =S\left(  x_{1}=L,x_{2}=L,\omega\right)  $.
\begin{figure}[ptb]
\begin{center}
\includegraphics[
height=5.77cm,
width=7.99cm
]{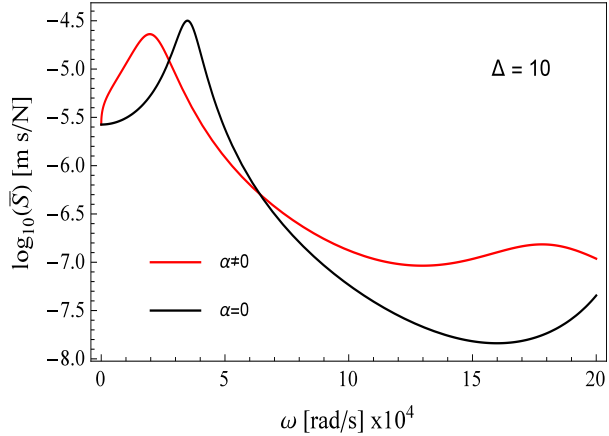}
\end{center}
\caption{The influence of the velocity-diffusive term on the cantilever
thermal power spectrum, for $\Delta=2d/B=10$.}%
\label{Figure 3}%
\end{figure}

\begin{center}%
\begin{tabular}
[c]{|c|c|}\hline
$\text{Quantity}$ & $\text{Value}$\\\hline
$L$ & $232.4~\mathrm{\mu m}$\\\hline
$B$ & $20.11~\mathrm{\mu m}$\\\hline
$H$ & $0.573~\mathrm{\mu m}$\\\hline
$\text{Young Modulus}$ & $3.92\times10^{11}~\mathrm{Pa}$\\\hline
$\text{Air density}$ & $1.2~\mathrm{kg/m}^{3}$\\\hline
$\text{Air -dynamic viscosity}$ & $1.83\times10^{-5}~\mathrm{Pas}$\\\hline
$\text{Water density}$ & $1000~\mathrm{kg/m}^{3}$\\\hline
$\text{Water-dynamic viscosity}$ & $1.0\times10^{-3}~\mathrm{Pas}$\\\hline
\end{tabular}

{\small Tab.2 - The geometrical and physical quantities utilized to carry out
the analysis.}
\end{center}

Figure \ref{Figure 2} shows the PSD of the cantilever tip deflection $\bar
{S}\left(  \omega\right)  =S\left(  \omega\right)  /2k_{B}T$, when the
dimensionless distance $\Delta=2d/B$ of the cantilever from the substrate is
constant and equal to $\Delta=10$ ($d$ is the distance in $\mu m$), for a
cantilever thermally excited in water (red curve) and in air (black curve).
Observe the large peak shifting towards the low frequency range when the
system is in water. At a first sight one may be surprised to see that the
heights of the two peaks for water and air are not significantly different.
However this can be easily explained considering that, although the imaginary
part of the complex compliance is larger in air if compared to water, the peak
frequency is about 3 times larger in air than in water. Hence, these two
opposite effects partially balance each-other when the quantity $\bar
{S}\left(  \omega\right)  $ is calculated. Beside this, one should also
consider that, even in air, the presence of the wall makes non negligible the
viscous dissipation related to the term $c\left(  x\right)  $ in
Eq.\ref{G_function}. Thus, a blunted resonant peak is expected to be observed
as indeed shown in Fig. \ref{Figure 2}. \begin{figure}[ptb]
\begin{center}
\includegraphics[
height=5.77cm,
width=7.99cm
]{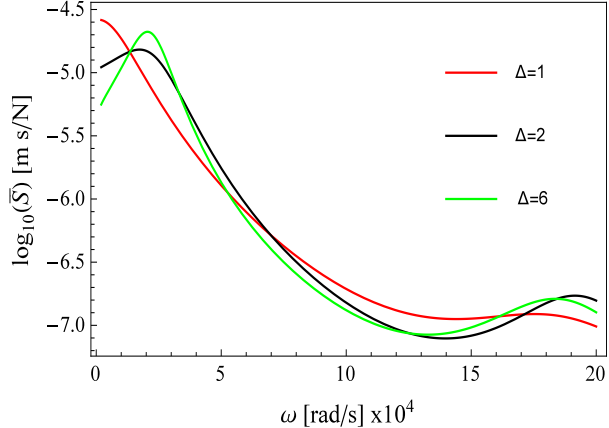}
\end{center}
\caption{The thermal power spectrum for different values of the dimensionless
distance $\Delta=2d/B$.}%
\label{Figure 4}%
\end{figure}In Fig.\ref{Figure 3}, the thermal power spectrum $\bar{S}\left(
\omega\right)  $ is shown, for $\alpha\neq0$ as results from numerical
fitting, and for $\alpha=0$. Fig. \ref{Figure 3} shows the very large
influence of the velocity-diffusive parameter in terms of peak resonance
position and the Q-factor values. This means that neglecting such effect can
lead to a strong overestimation of the first resonance frequency and also to
strongly underestimate the noise affecting the dAFM at larger frequencies.
This error in the evaluation of the cantilever thermal response to Brownian
forcing, can in turn lead to a wrong calibration of the instrument. In
particular, a bad estimation of the cantilever resonances and damping
properties, yields an invalid estimation of the system properties and
therefore may result in a non-correct operation of the instrument. This means
that our drag model can be an extremely useful tool in the calibration process
of a dAFM.

\begin{figure}[ptb]
\begin{center}
\includegraphics[
width=13.0cm
]{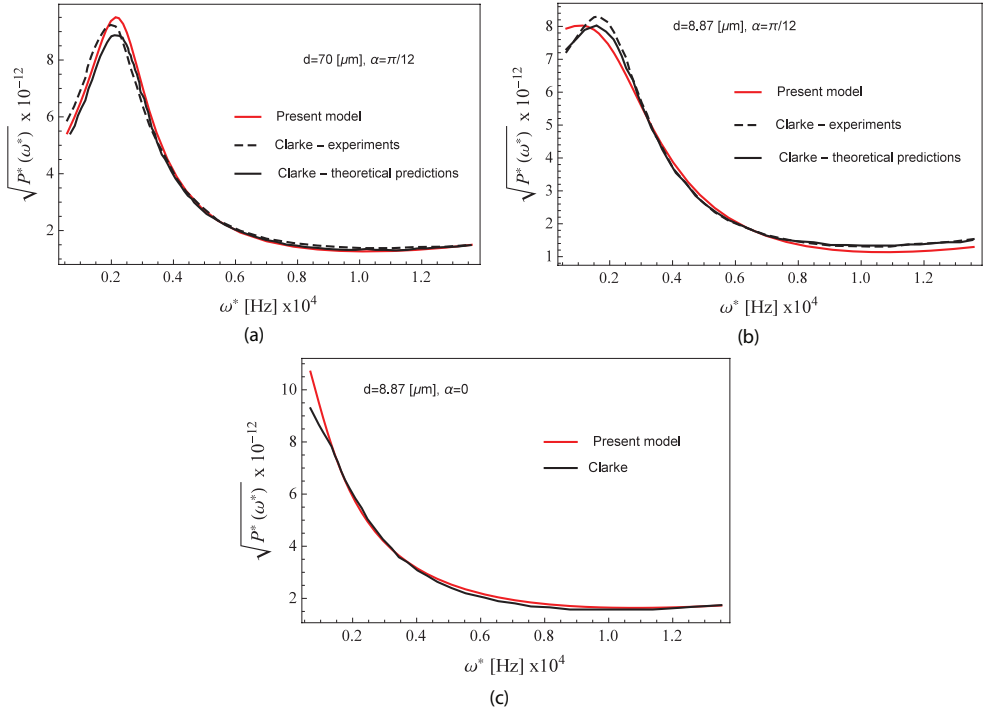}
\end{center}
\caption{The thermal power spectrum $\sqrt{P^{\ast}\left(  \omega^{\ast
}\right)  }$ in water near a wall. Comparison between experiments (dashed
black curve) and theoretical predictions (solid black curve) presented in
Ref.\cite{14} with our analytical model (solid red curve) for a tilt angle
$\pi/12$ to the wall and separations $d=70\mu m$ (a) and $d=8.87\mu m$ (b).
Theoretical predictions from Ref.\cite{14} (solid black curve) are also
compared with our analytical model (solid red curve) for a tilt $\alpha=0$ and
separation $d=8.87\mu m$ (c).}%
\label{Figure 5}%
\end{figure}In Fig. \ref{Figure 4} the cantilever thermal power spectrum
$\bar{S}\left(  \omega\right)  $ at different distances from the substrate is
shown. Three beam-wall distances are considered. The figure shows that the
closer the cantilever is to the wall, the more the spectrum is shifted toward
lower frequencies, whereas the second resonance peak occurs at higher
frequencies. In particular, for the lowest value of $\Delta$ here considered,
i.e. $\Delta=1$, the first resonant peaks almost disappears. The reason for
such a strong influence of the substrate distance should be sought in the
change of the viscous response of the liquid. Indeed at small beam-wall
distances the viscous coefficient $c$ increases with $d^{-3}$ as the distance
$d$ is reduced.

Considering that in many applications the cantilever is slightly tilted along
its length, the effect of the $x$-dependent distance of the beam cross section
from the substrate should be accurately taken into account to correctly
estimate the cantilever response. This is the case of the experimental data
shown in Ref.\cite{14}, where a cantilever in water is moved close to the
substrate, with a tilt $\alpha=\pi/12$ (geometrical and physical quantities
are the same listed in Table 2), and the angular oscillations are measured. In
order to compare our analytical model with the numerical and experimental
thermal power spectra presented in \cite{14}, we have numerically calculated
the interpolating functions of the three coefficients of the function
$G\left(  x,t\right)  $ (Eq.\ref{G_function}), i.e. $c\left(  x\right)  $,
$m\left(  x\right)  $, and $\alpha\left(  x\right)  $, by means of the values
reported in Table 1. Then we have calculated the PSD $S\left(  \omega\right)
$ (see Eq.\ref{tot_CPSD}) given by both the vertical displacement and the
angular rotation of the cantilever free end, and finally we have defined
$P^{\ast}\left(  \omega^{\ast}\right)  =\left(  \bar{A}^{2}/\omega_{0}\right)
S\left(  \omega\right)  $, where $\bar{A}=\sqrt{k_{B}TL^{3}/\left(  EJ\right)
}$, $\omega^{\ast}=\omega_{0}\omega$, and $\omega_{0}=\sqrt{EJ/\left(  \rho
AL^{4}\right)  }$. In Figure \ref{Figure 5} we show the comparison of the
quantity $\sqrt{P^{\ast}\left(  \omega^{\ast}\right)  }$ calculated by
employing our analytical model (with $c_{1}=0$ and $c_{2}=0.86$ in
Eq.\ref{tot_CPSD}) and the experimental data and theoretical predictions
presented in \cite{14}, conducted in water, using a molecular force probe with
pyramidal tip. In Fig.\ref{Figure 5}-a the cantilever is tilted at $\pi/12$ to
the wall with a separation $d=70\mu m$, in Fig.\ref{Figure 5}-b the distance
is reduced to $d=8.87\mu m$ (same tilt), in Fig. \ref{Figure 5}-c the
cantilever has no tilt ($\alpha=0$) and the separation is $d=8.87\mu m$. In
this last case no experimental data are presented in Ref. \cite{14} and we
show the comparison between our analytical model and the one utilized in
\cite{14}, where the fluid - cantilever interaction is described considering
the Stokes' expression for the drag on an oscillating circular cylinder. It is
possible to observe a good agreement between our results (solid red lines),
the experimental data (dashed black lines) and theoretical predictions (solid
black lines) reported Ref.\cite{14}. It is worth of being mentioned that our
analytical model is able to fit very well the experimental data (see Fig.
\ref{Figure 5}-a). When the tilted cantilever is much closer to the wall (i.e.
for $\Delta<1$) we expect to have a smaller degree of agreement, confirmed in
Fig. \ref{Figure 5}-b, for the data presented in literature showed results for
only a couple values of $\Delta$. In particular, considering the near wall
case, we have found data only for $\Delta=0.5,1$ (see Table 1) that are not
sufficient for an accurate estimation of the damping coefficient
$c$.\begin{figure}[ptb]
\begin{center}
\includegraphics[
width=9cm
]{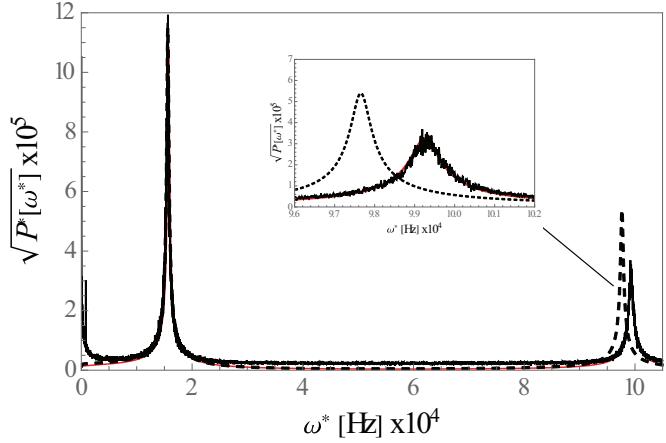}
\end{center}
\caption{Our experimental thermal power spectrum $\sqrt{P^{\ast}\left(
\omega^{\ast}\right)  }$ in air (solid black curve), far from the wall,
compared with our theoretical predictions (solid red curve) and the
theoretical results presented in Ref.\cite{14} (dashed black curve). Our
analytical model perfectly fit the experimental curve, also at the second
resonant peak, as shown in the inset.}%
\label{Figure 6}%
\end{figure}To completely assess our analytical model, we have carried out
experiments with an AFM NT-MDT Ntegra Aura (Tribolab, Politecnico di Bari,
Bari, Italy) on a CSG01 rectangular silicon cantilever with tip, and
dimensions $\left(  L,B,H\right)  =\left(  350,30,1\right)  \mathrm{\mu m}$.
The thermal oscillations of the free-end of the cantilever have been acquired,
in the frequency range $0<\omega^{\ast}<105~\mathrm{kHz}$, where the first and
the second resonant peaks of the beam are present. In Figure \ref{Figure 6}
our experimentally measured thermal power spectrum$\sqrt{P^{\ast}\left(
\omega^{\ast}\right)  }$ is compared with our theoretically predicted
response. An almost perfect agreement between the experimental data and our
theoretical model is obtained. The inset shows, in particular, that this
almost perfect matching still holds true for the second resonant peak,\ a
condition not easy achievable with other techniques (see dashed black curve in
Fig.\ref{Figure 6} corresponding to the model reported in Ref.\cite{14}). Our
results definitively show that the presented analytical model is able to
accurately predict the dynamics of a dAFM cantilever in a wide frequency
range, thus making it a possible tool for calibration procedures and high
performance measurements.

\section{Conclusions}

The present study is concerned with the dAFM cantilever dynamics. More
specifically, the attention has been focused on the interactions of the
cantilever with the surrounding fluid. The impacts of the fluid molecules on
the beam, indeed, generate the so called Brownian thermal noise, which is
related to the macroscopic linear response of the system. An analytical model
of the fluid-structure interactions has been presented, which takes into
account of inertial, damping, and diffusive terms. In particular, the force
the liquid exerts on the body has been heuristically derived, and it is
described as the sum of three contributions, which have been evaluated through
the best fitting of accurate computational fluid dynamics (adimensionalized)
data of a 2D fluid flowing around an oscillating rectangular cross section,
found in the literature. Experiments have been also carried out, and the
thermal oscillations of a dAFM cantilever operating in air have been acquired.
The experimental data have been perfectly fitted by means of our analytic
response of the beam, in a wide frequency range. The results presented in this
paper clearly demonstrate that our analytical model is an extremely useful and
accurate tool to predict the cantilever dynamics, and therefore could be
utilized to calculate the cantilever spring constants in the common
calibration procedures, thus improving the quality of the measurements.

\appendix\textbf{A. The Susceptibility Function}

Here we calculate the complex compliance $\chi\left(  x_{1},x_{2}%
,\omega\right)  $, which is the solution of the motion equation%
\begin{equation}
\frac{\partial^{4}\chi\left(  x_{1},x_{2},\omega\right)  }{\partial x^{4}%
}-B\left(  x,\omega\right)  ^{4}\chi\left(  x_{1},x_{2},\omega\right)
=\delta\left(  x_{1}-x_{2}\right)  \label{governing_eq1}%
\end{equation}
that can be solved by considering the following equivalent problem%
\begin{equation}
\frac{\partial^{4}\chi\left(  x_{1},x_{2},\omega\right)  }{\partial x^{4}%
}-B\left(  x,\omega\right)  ^{4}\chi\left(  x_{1},x_{2},\omega\right)  =0
\label{governing_eq_2}%
\end{equation}
with the boundary conditions%
\begin{align}
\chi^{I}\left(  0,x_{2},\omega\right)   &  =\left.  \frac{\partial\chi
^{I}\left(  x_{1},x_{2},\omega\right)  }{\partial x_{1}}\right\vert _{x_{1}%
=0}=0\label{boundary2}\\
\left[  \chi\left(  x_{1},x_{2},\omega\right)  \right]  _{x_{1}=x_{2}^{-}%
}^{x_{1}=x_{2}^{+}}  &  =\left[  \frac{\partial\chi\left(  x_{1},x_{2}%
,\omega\right)  }{\partial x_{1}}\right]  _{x_{1}=x_{2}^{-}}^{x_{1}=x_{2}^{+}%
}=\left[  \frac{\partial^{2}\chi\left(  x_{1},x_{2},\omega\right)  }%
{\partial^{2}x_{1}}\right]  _{x_{1}=x_{2}^{-}}^{x_{1}=x_{2}^{+}}=0\nonumber\\
\left[  \frac{\partial^{3}\chi\left(  x_{1},x_{2},\omega\right)  }%
{\partial^{3}x_{1}}\right]  _{x_{1}=x_{2}^{-}}^{x_{1}=x_{2}^{+}}  &
=1\nonumber\\
\left.  \frac{\partial^{2}\chi^{II}\left(  x_{1},x_{2},\omega\right)
}{\partial^{2}x_{1}}\right\vert _{x_{1}=1}  &  =\left.  \frac{\partial^{3}%
\chi^{II}\left(  x_{1},x_{2},\omega\right)  }{\partial^{3}x_{1}}\right\vert
_{x_{1}=1}=0\nonumber
\end{align}
where the third condition can be derived by integrating Eq.\ref{governing_eq1}%
\begin{equation}
\int_{x_{2}-\varepsilon}^{x_{2}+\varepsilon}\frac{\partial^{4}\chi\left(
x_{1},x_{2},\omega\right)  }{\partial x_{1}^{4}}dx_{1}-B\left(  x_{1}%
,\omega\right)  ^{4}\int_{x_{2}-\varepsilon}^{x_{2}+\varepsilon}\chi\left(
x_{1},x_{2},\omega\right)  dx_{1}=1 \label{third_bc}%
\end{equation}
which for $\varepsilon\rightarrow0^{+}$ becomes $\left[  \partial^{3}%
\chi\left(  x_{1},x_{2},\omega\right)  /\partial^{3}x_{1}\right]
_{x_{1}=x_{2}^{-}}^{x_{1}=x_{2}^{+}}=1$. This condition, in particular, shows
that the third derivative of the susceptibility function $\chi\left(
x_{1},x_{2},\omega\right)  $ is discontinue in $x_{2}$, thus requiring to
define two different functions $\chi^{I}\left(  x_{1},x_{2},\omega\right)  $
in the interval $0<x_{1}<x_{2}$, and $\chi^{II}\left(  x_{1},x_{2}%
,\omega\right)  $ for $x_{2}<x_{1}<1$. The second boundary condition in
Eqs.\ref{boundary2}, represents a continuity condition for the susceptibility
function $\chi\left(  x_{1},x_{2},\omega\right)  $ and its first and second
derivatives. The general integral of Eq. \ref{governing_eq_2} is%
\begin{align}
\chi^{i}\left(  x_{1},x_{2},\omega\right)   &  =a_{1i}\cos\left[  B\left(
x_{1},\omega\right)  x_{1}\right]  +a_{2i}\sin\left[  B\left(  x_{1}%
,\omega\right)  x_{1}\right]  +\label{generic_solution}\\
&  +a_{3i}\cosh\left[  B\left(  x_{1},\omega\right)  x_{1}\right]
+a_{4i}\sinh\left[  B\left(  x_{1},\omega\right)  x_{1}\right] \nonumber
\end{align}
for $i=I,II$. The eight coefficients $a_{1i}=a_{1i}\left(  x_{2},B\right)  $,
and hence the complete solution to the Eq.\ref{governing_eq_2}, can be
calculated by means of the boundary conditions Eqs.\ref{boundary2}, for
$i=I,II$, and making use of the symmetry property $\chi\left(  x_{1}%
,x_{2},\omega\right)  =\chi\left(  x_{2},x_{1},\omega\right)  $. For example,
in the interval $0<x_{1}<x_{2}$%
\begin{align}
\chi^{I}\left(  x_{1},x_{2},\omega\right)   &  =\frac{1}{4B^{3}\left(  1+\cos
B\cosh B\right)  }\times\{\label{Susc_func}\\
&  \left[  \cosh\left(  Bx_{1}\right)  -\cos\left(  Bx_{1}\right)  \right]
\nonumber\\
&  [\sin\left(  Bx_{2}\right)  +\sinh\left(  Bx_{2}\right)  \left(  1+\cos
B\cosh B-\sin B\sinh B\right) \nonumber\\
&  \cosh\left(  Bx_{2}\right)  \left(  \sin B\cosh B-\cos B\sinh B\right)
-\sin\left[  B\left(  1-x_{2}\right)  \right]  \cosh B\nonumber\\
&  +\cos\left[  B\left(  1-x_{2}\right)  \right]  \sinh B]+\nonumber\\
&  \left[  \sin\left(  Bx_{1}\right)  -\sinh\left(  Bx_{1}\right)  \right]
\nonumber\\
&  [\cos\left(  Bx_{2}\right)  +\cosh\left(  Bx_{2}\right)  \left(  1+\sin
B\sinh B+\cos B\cosh B\right) \nonumber\\
&  -\sinh\left(  Bx_{2}\right)  \left(  \sin B\cosh B+\cos B\sinh B\right)
-\sin\left[  B\left(  1-x_{2}\right)  \right]  \sinh B\nonumber\\
&  +\cos\left[  B\left(  1-x_{2}\right)  \right]  \cosh B]\}\nonumber
\end{align}
.

\appendix\textbf{B.The Fluctuation Dissipation Theorem for a 1DOF system}

The Fluctuation Dissipation Theorem (FDT) for a single degree of freedom
system is here derived. For this purpose, we consider a simple mass $M$ which
oscillates along a single direction with the law $u\left(  t\right)  $,
excited by an external stochastic force $F\left(  t\right)  $, and connected
to a rigid frame through both a spring with elastic constant $k$ and a general
linear dissipative element identified by the linear response function
$L\left(  t\right)  $. The motion of the mass $M$ is governed by the following
differential equation%
\begin{equation}
M\ \ddot{u}\left(  t\right)  +\int_{0}^{t}L\left(  t-\tau\right)  \ \dot
{u}\left(  \tau\right)  d\tau+k\ u\left(  t\right)  =F\left(  t\right)
\label{sdof_motion_eq}%
\end{equation}
where we assumed that the fluctuating force has been switched on at time
$t=0$. We multiply the previous equation by $u\left(  0\right)  $ and we
calculate the main value
\begin{equation}
M\ \langle\ddot{u}\left(  t\right)  u\left(  0\right)  \rangle+\int_{0}%
^{t}L\left(  t-\tau\right)  \ \langle\dot{u}\left(  \tau\right)  u\left(
0\right)  \rangle d\tau+k\ \langle u\left(  t\right)  u\left(  0\right)
\rangle=\langle F\left(  t\right)  u\left(  0\right)  \rangle
\label{Eq_mean_val}%
\end{equation}
The equation then becomes%
\begin{equation}
M\ \ddot{R}\left(  t\right)  +\int_{0}^{t}L\left(  t-\tau\right)  \ \dot
{R}\left(  \tau\right)  d\tau+k\ R\left(  t\right)  =0
\label{sdof_correl_motion_Eq}%
\end{equation}
where we have defined the correlation function $R\left(  t\right)  =\langle
u\left(  t\right)  u\left(  0\right)  \rangle$, and the term $\langle F\left(
t\right)  u\left(  0\right)  \rangle=0$. We consider the variable change
$X\left(  t\right)  +R_{0}=R\left(  t\right)  $, being $R_{0}=\langle
u^{2}\rangle$%
\begin{equation}
M\ \ddot{X}\left(  t\right)  +\int_{0}^{t}L\left(  t-\tau\right)  \ \dot
{X}\left(  \tau\right)  d\tau+k\ X\left(  t\right)  =-H\left(  t\right)
kR_{0} \label{Xeq}%
\end{equation}
Now recall that the system response can be written using its response function
$\chi\left(  t\right)  $, which is the displacement caused by an impulsive
unit force placed at $t=0$, i.e. by a dirac delta force $\delta\left(
t\right)  $. Then exploiting the linearity one can write%
\begin{equation}
X\left(  t\right)  =\int_{-\infty}^{t}\chi\left(  t-\tau\right)  \left[
-H\left(  t\right)  kR_{0}\right]  d\tau=-kR_{0}\int_{0}^{t}\chi\left(
t-\tau\right)  d\tau=-kR_{0}\int_{0}^{t}\chi\left(  \tau\right)
d\tau\label{Xsol}%
\end{equation}
Taking the time derivative of Eq. (\ref{Xsol}) we obtain the Fluctuation
Dissipation Theorem%
\begin{equation}
\chi\left(  t\right)  =-\frac{1}{k_{B}T}H\left(  t\right)  \langle u\left(
0\right)  \dot{u}\left(  t\right)  \rangle\label{FDT_1dof}%
\end{equation}
where we have used the equipartition theorem \cite{21}, which, in this case,
simply states that $kR_{0}=\langle ku^{2}\rangle=k_{B}T$, where $k_{B}$ is the
Boltzmann constant. Recalling that $\int dtH\left(  t\right)  e^{-i\omega
t}=\wp\frac{1}{i\omega}+\pi\delta\left(  \omega\right)  $, where
$\delta\left(  \omega\right)  $ is the Dirac delta function, and Fourier
transforming the Eq. \ref{FDT_1dof} leads%
\begin{equation}
\chi\left(  \omega\right)  =-\frac{1}{2k_{B}T}\frac{1}{\pi}\int_{-\infty
}^{+\infty}\frac{\omega^{\prime}R\left(  \omega^{\prime}\right)  }%
{\omega-\omega^{\prime}}d\omega^{\prime}-\frac{1}{2k_{B}T}i\omega R\left(
\omega\right)  \label{Fourier_FDT_1dof}%
\end{equation}
from which it follows the spectral version of the Fluctuation Dissipation
theorem
\begin{equation}
R\left(  \omega\right)  =-2k_{B}T\frac{\operatorname{Im}\left[  \chi\left(
\omega\right)  \right]  }{\omega} \label{FDT_1dof_freq}%
\end{equation}
and%
\begin{equation}
\operatorname{Re}\left[  \chi\left(  \omega\right)  \right]  =\frac{1}{\pi}%
\wp\int_{-\infty}^{+\infty}\frac{\operatorname{Im}\left[  \chi\left(
\omega^{\prime}\right)  \right]  }{\omega-\omega^{\prime}}d\omega^{\prime}
\label{KK1}%
\end{equation}
which also gives%
\begin{equation}
\operatorname{Im}\left[  \chi\left(  \omega\right)  \right]  =-\frac{1}{\pi
}\wp\int_{-\infty}^{+\infty}\frac{\operatorname{Re}\left[  \chi\left(
\omega^{\prime}\right)  \right]  }{\omega-\omega^{\prime}}d\omega^{\prime}
\label{KK2}%
\end{equation}
Eqs. (\ref{KK1},\ref{KK2}) are the well known Kramer-Kronig relations.

\appendix\textbf{C.The Fluctuation Dissipation Theorem for the cantilever
case}

In this paper we study a continuos body, the cantilever, subjected to thermal
driven fluctuations. Therefore in this section we will particularize the FDT
for this special case. Recall Eq. (\ref{complete_equation}), which reads%

\begin{align}
&  EJ\frac{\partial^{4}w\left(  x_{2},t\right)  }{\partial x_{2}^{4}}+\left[
\rho A+\mu\left(  x_{2}\right)  \right]  \frac{\partial^{2}w\left(
x_{2},t\right)  }{\partial t^{2}}+c\left(  x_{2}\right)  \frac{\partial
w\left(  x_{2},t\right)  }{\partial t}+\label{original eq}\\
+\alpha\left(  x_{2}\right)  \int_{0}^{t}\frac{1}{\sqrt{t-\tau}}\frac
{\partial^{2}w\left(  x_{2},\tau\right)  }{\partial\tau^{2}}d\tau &
=f_{B}\left(  x_{2},t\right) \nonumber
\end{align}
To derive the Fluctuation Dissipation theorem in this case we need to
calculate the motion equation for the correlation function%
\begin{equation}
R\left(  x_{1},x_{2},t\right)  =\left\langle w\left(  x_{1},0\right)  w\left(
x_{2},t\right)  \right\rangle \label{Corr_func}%
\end{equation}
so multiplying Eq. (\ref{original eq}) times $w\left(  x_{1},0\right)  $,
taking the ensemble average and recalling that Brownian force $f_{B}\left(
x_{2},t\right)  $ is independent of $w\left(  x_{1},0\right)  $, i.e.
$\left\langle w\left(  x_{1},0\right)  f_{B}\left(  x_{2},t\right)
\right\rangle =0$ we obtain%
\begin{align}
&  EJ\frac{\partial^{4}R\left(  x_{1},x_{2},t\right)  }{\partial x_{2}^{4}%
}+\left[  \rho A+\mu\left(  x_{2}\right)  \right]  \frac{\partial^{2}R\left(
x_{1},x_{2},t\right)  }{\partial t^{2}}+c\left(  x_{2}\right)  \frac{\partial
R\left(  x_{1},x_{2},t\right)  }{\partial t}+\label{Req}\\
+\alpha\left(  x_{2}\right)  \int_{0}^{t}\frac{1}{\sqrt{t-\tau}}\frac
{\partial^{2}R\left(  x_{1},x_{2},t\right)  }{\partial\tau^{2}}d\tau &
=0\nonumber
\end{align}
Eq. (\ref{Req}) is an homogeneous equation that we need to solve for $t>0$.
Now let us change the integration variable using the replacement $R\left(
x_{1},x_{2},t\right)  \rightarrow R\left(  x_{1},x_{2},0\right)  +X\left(
x_{1},x_{2},t\right)  $, to get%
\begin{align}
&  EJ\frac{\partial^{4}X\left(  x_{1},x_{2},t\right)  }{\partial x_{2}^{4}%
}+\left[  \rho A+\mu\left(  x_{2}\right)  \right]  \frac{\partial^{2}X\left(
x_{1},x_{2},t\right)  }{\partial t^{2}}+c\left(  x_{2}\right)  \frac{\partial
X\left(  x_{1},x_{2},t\right)  }{\partial t}+\label{XXeq}\\
+\alpha\left(  x_{2}\right)  \int_{0}^{t}\frac{1}{\sqrt{t-\tau}}\frac
{\partial^{2}X\left(  x_{1},x_{2},t\right)  }{\partial\tau^{2}}d\tau &
=-EJ\frac{\partial^{4}R\left(  x_{1},x_{2},0\right)  }{\partial x_{2}^{4}%
}H\left(  t\right) \nonumber
\end{align}
Using the response function $\chi\left(  x_{1},x_{2},t\right)  $
Eq.(\ref{XXeq}) can be solved to give
\begin{equation}
X\left(  x_{1},x_{2},t\right)  =-EJ\int d\xi\int_{0}^{t}d\tau\chi\left(
x_{1},\xi,\tau\right)  \frac{\partial^{4}R\left(  x_{1},\xi,0\right)
}{\partial\xi^{4}} \label{sol}%
\end{equation}
using the continuous version of the equipartition theorem (see Appendix D)%
\begin{equation}
EJ\frac{\partial^{4}R\left(  x_{1},x_{2},0\right)  }{\partial x_{2}^{4}%
}H\left(  t\right)  =k_{B}T\delta\left(  x_{1}-x_{2}\right)
\label{equiripartition}%
\end{equation}
where $\delta\left(  x\right)  $ is the Dirac delta function, and taking the
time derivative we obtain the fluctuation dissipation theorem for our
cantilever
\begin{equation}
\frac{\partial R\left(  x_{1},x_{2},t\right)  }{\partial t}H\left(  t\right)
=-k_{B}T\chi\left(  x_{2},x_{1},t\right)  \label{FDT continuous}%
\end{equation}
and taking the Fourier transform%
\begin{equation}
R\left(  x_{1},x_{2},\omega\right)  =-2k_{B}T\frac{\operatorname{Im}%
\chi\left(  x_{1},x_{2},\omega\right)  }{\omega} \label{FDT cantilever}%
\end{equation}
and%
\begin{align}
\operatorname{Re}\chi\left(  x_{1},x_{2},\omega\right)   &  =\frac{1}{\pi}%
\wp\int_{-\infty}^{+\infty}\frac{\operatorname{Im}\chi\left(  x_{1}%
,x_{2},\omega^{\prime}\right)  }{\omega-\omega^{\prime}}d\omega^{\prime
}\label{Re_Im_parts}\\
\operatorname{Im}\chi\left(  x_{1},x_{2},\omega\right)   &  =-\frac{1}{\pi}%
\wp\int_{-\infty}^{+\infty}\frac{\operatorname{Re}\chi\left(  x_{1}%
,x_{2},\omega^{\prime}\right)  }{\omega-\omega^{\prime}}d\omega^{\prime
}\nonumber
\end{align}

\appendix\textbf{D.Equipartition theorem for the cantilever case}

Let us calculate the potential energy $U$, i.e. the elastic energy, of our
cantilever. We observe that the potential energy only depends on the
configuration of the system, where the displacement field $w\left(  x\right)
$ must satisfy the equation
\begin{equation}
EJ\frac{\partial^{4}w\left(  x\right)  }{\partial x^{4}}=f\left(  x\right)
\label{Equil_eq}%
\end{equation}
where $f\left(  x\right)  $ is the sum of the forces acting on the cantilever
plus the inertia forces. Then, because of linearity, we write%
\begin{equation}
U=\frac{1}{2}\int dxf\left(  x\right)  w\left(  x\right)  =\frac{1}{2}EJ\int
dx\frac{\partial^{4}w\left(  x\right)  }{\partial x^{4}}w\left(  x\right)
\label{Potential_energy}%
\end{equation}
Now let us calculate the first variation of the potential energy as%
\begin{equation}
\delta U=EJ\int dx\frac{\partial^{4}w\left(  x\right)  }{\partial x^{4}}\delta
w\left(  x\right)  \label{energy variation}%
\end{equation}
where, by integrating by parts, we have used%
\begin{equation}
\int dx\frac{\partial^{4}\delta w\left(  x\right)  }{\partial x^{4}}w\left(
x\right)  =\int dx\frac{\partial^{4}w\left(  x\right)  }{\partial x^{4}}\delta
w\left(  x\right)  \label{integrate by part}%
\end{equation}
In order to exploit the equipartition theorem let us first discretize Eq.
(\ref{energy variation}) as
\begin{equation}
\delta U=\sum_{k}EJw_{k}^{\left(  4\right)  }\delta w_{k}\delta x
\label{discretized}%
\end{equation}
where $w_{k}=w\left(  x_{k}\right)  $ and $w_{k}^{\left(  4\right)  }\ $is the
fourth derivative of $w$ calculate at $x=x_{k}$. From Eq. (\ref{discretized})
we have%
\begin{equation}
\frac{\delta U}{\delta w_{k}}=EJw_{k}^{\left(  4\right)  }\delta x
\label{eq11}%
\end{equation}
Using the equipartition theorem, where now the Lagrangian coordinates are the
quantities $w_{k}$, we get%
\begin{equation}
k_{B}T\delta_{hk}=\left\langle w_{h}\frac{\delta U}{\delta w_{k}}\right\rangle
=EJ\left\langle w_{h}w_{k}^{\left(  4\right)  }\right\rangle \delta x
\label{eq22}%
\end{equation}
where $\delta_{hk}$ is the Kronecker delta, then recalling that $\left\langle
w_{h}w_{k}^{\left(  4\right)  }\right\rangle =\partial^{4}R\left(  x_{h}%
,x_{k},0\right)  /\partial x_{k}^{4}$ we get%
\begin{equation}
EJ\frac{\partial^{4}R\left(  x_{h},x_{k},0\right)  }{\partial x_{k}^{4}}%
=k_{B}T\delta\left(  x_{k}-x_{h}\right)  \label{eq33}%
\end{equation}
where we have used that in the continuum limit $\delta_{hk}/\delta
x\rightarrow\delta\left(  x_{k}-x_{h}\right)  $.

\end{document}